\documentclass[pdftex,twocolumn,epjc3]{svjour3}          % twocolumn

\RequirePackage[T1]{fontenc}

\smartqed  % flush right qed marks, e.g. at end of proof

\RequirePackage{graphicx}
\RequirePackage{mathptmx}      % use Times fonts if available on your TeX system
\RequirePackage{flushend}
\RequirePackage[numbers,sort&compress]{natbib}
\RequirePackage[colorlinks,citecolor=blue,urlcolor=blue,linkcolor=blue]{hyperref}
\journalname{Eur. Phys. J. C}

\usepackage{amsmath}
\usepackage{subfiles}
\usepackage{caption}
\usepackage{subcaption}
\begin{document}
\emergencystretch 3em
%\tableofcontents

\title{Data-Directed Search for New Physics based on Symmetries of the SM}
  
\author{
  Mattias Birman\thanksref{addr1,e1}
  \and
  Benjamin Nachman\thanksref{addr2,addr3}
  \and
  Raphael Sebbah\thanksref{addr1}
  \and
  Gal Sela\thanksref{addr1}   
  \and  
  Ophir Turetz\thanksref{addr1}
  \and
  Shikma Bressler\thanksref{addr1}
}

\thankstext{e1}{Corresponding author: mattias.birman@weizmann.ac.il}

\institute{Department of Particle Physics \& Astrophysics, Weizmann Institute of Science, Rehovot, Israel\label{addr1}
\and
Physics Division, Lawrence Berkeley National Laboratory, Berkeley, CA 94720, USA\label{addr2}
\and
Berkeley Institute for Data Science, University of California, Berkeley, CA 94720, USA\label{addr3}}
 
\date{Received: 22 March 2022 / Accepted: 20 May 2022}

\maketitle
 
\begin{abstract}
We propose exploiting symmetries (exact or approximate) of the Standard Model (SM) to search for physics Beyond the Standard Model (BSM) using the data-directed paradigm (DDP). Symmetries are very powerful because they provide two samples that can be compared without requiring simulation. Focusing on the data, exclusive selections which exhibit significant asymmetry can be identified efficiently and marked for further study.
Using a simple and generic test statistic which compares two matrices
already provides good sensitivity, only slightly worse than that of
the profile likelihood ratio test statistic which relies on the exact
knowledge of the signal shape. This can be exploited for rapidly
scanning large portions of the measured data, in an attempt to
identify regions of interest. We also demonstrate that weakly supervised
Neural Networks could be used for this purpose as well. 
\end{abstract}

\section{Introduction}
\label{sec:intro} 
Despite its success in describing the elementary particles and their interactions, the Standard Model (SM) is still incomplete, e.g., it does not account for neutrino masses, baryon asymmetry or dark matter. Thus, the discovery of physics beyond the Standard Model (BSM) is one of the main goals of particle physics. In particular, it is a core component of the physics program of the two multipurpose experiments, ATLAS \cite{atlas} and CMS \cite{cms} at the Large Hadron Collider (LHC) at CERN. 
So far hundreds of searches for BSM physics have been conducted, but no significant deviation from
the SM predictions has been found. With only a few exceptions (e.g., \cite{D0:2000vuh,H1:2008aak,gensearchCDF,gensearch1,gensearch2,gensearch3}), most of
these searches were conducted following the blind-analysis paradigm according to which the data is only looked at in the last step of the analysis, after most of the time and efforts were invested. Moreover, since the data are looked at in the end, these analyses  
were typically designed to inspect a specific region of the
\emph{observables space} - the space spanned by all observables of the
recorded data. 
%This is due to the "blind-analysis" paradigm, % where
% searches are mainly performed based on theoretically motivated BSM
% scenarios,
%according to which the tested regions are defined at an early stage of the
%analysis, and the data is only looked at in the end,
%after most of the efforts 
%and time were invested. 
As a result, despite thousands of person-years invested, a large portion of the observables
space has yet to be fully exploited (see also Ref.~\cite{Craig:2016rqv,Kim:2019rhy}).
The risk of missing a discovery by studying only a limited number of final states could be mitigated by prioritizing the searches and focusing the efforts on high priority ones. Traditionally, this is mostly done based on theoretical considerations. However, by now, the searches with the strongest theoretical motivation have mostly been conducted and to a large extent, none of the many remaining ones are, a priori, more motivated than the others. This calls for investigating additional search paradigms.

Complementary to the blind  searches, we propose 
extending the discovery potential of the LHC 
with a data-directed paradigm
(DDP). Similarly to \cite{gensearch1,gensearch3,gensearchCDF}, its principal objective is to efficiently scan large portions
of the observables space for hints of new physics (NP), but unlike \cite{gensearch1,gensearch3,gensearchCDF} without using any Monte-Carlo (MC) simulation. We look directly at the data, in an attempt to identify regions 
in the observables space that exhibit deviations from a theoretically well established property of the SM. Such regions should be considered as data-directed BSM
hypotheses, as opposed to theoretically-motivated ones, and
could be studied using traditional data-analysis methods. 
As detailed
in \cite{ddp-bumphunt}, a search in the DDP can be implemented with
two key ingredients: a) a theoretically well established property of
the SM and b) an efficient algorithm to search for deviations from
this property. \\

In this work, we show that any symmetry of the SM can be exploited in such a
data-directed search. Symmetries can be used to split the data into two 
mutually exclusive samples which should only differ by statistical
fluctuations. By comparing them, we become sensitive to any potential
BSM process which breaks this symmetry. In some cases, systematic detector effects could also affect the symmetry.  There are methods to account for these effects in principle and so we do not consider them further in this proof-of-principle study. In an experimental realization of the symmetry-based DDP search, such systematic effects must be taken into account.

The concept of exploiting symmetries of the SM for data-driven BSM searches was previously proposed in \cite{emmePheno} and \cite{chris3}. It is also
implemented in the ATLAS search for lepton flavor violating (LFV)
decays of the Higgs (H) Boson \cite{emmeRun1}, and the search for asymmetry between $e^{+}\mu^{-}$ and $e^{-}\mu^{+}$ events \cite{epmum}. In the former, the SM background is estimated from the data using the
electron-muon ($e/\mu$) symmetry method, based on the premise
that kinematic properties of SM processes are, to a good approximation,
symmetric under the exchange of prompt electrons and prompt muons\footnote{This approximate
symmetry derives from the \emph{lepton flavor universality} of the electroweak
force. Phase-space effects and Higgs interactions only violate it at negligible levels within the energy range of LHC collisions, since the mass of the two leptons is negligible.}. In this case, the
sample of all recorded data events with one
electron and one muon in the final state is split into the $e\mu$ and $\mu e$
samples, which differ only by the $p_\text{T}$ ordering of the two
leptons. The Higgs LFV signal is expected to contribute only to one of these two samples, while the other is used
as the background estimate. In \cite{emmeRun1}, it was shown that systematic effects which violate the expected SM symmetry, e.g., the different detection efficiencies of electrons and muons, can be accounted for and the symmetry can be restored\footnote{This effect is suppressed in \cite{epmum} since the lepton (electron or muon) detection efficiencies in ATLAS depend on the lepton's $p_\text{T}$, but not on their charge.}.
However, the implementation of the search still follows the blind-analysis paradigm where only a specific signal is searched for, in a small theoretically-motivated
subset of the observables space.

In terms of the DDP proposed here, no specific signal is searched
for. Instead, the full $e\mu$ and $\mu e$ samples are compared in
many different sub-samples (corresponding to exclusive selections of the data),
and any significant deviation observed is considered a potential sign for NP, to be further
investigated. Thus, sensitivity to many more possible BSM
processes and scenarios is enabled, and this $e\mu/\mu e$ comparison becomes a general test for
lepton flavor universality in the final state containing one electron and one muon. Similarly, different final states including a number of 
electrons, muons and other objects can be probed ($ee$ vs
$\mu\mu$, $e+$jet vs $\mu+$jet, etc.), each
potentially sensitive to different BSM manifestations.
In this context, the recent hints for non-universality in the R$_K$ measurements from LHCb \cite{rk} are in fact hints of an asymmetry between the $ee$ and  $\mu\mu$ samples in the decay of $b$ hadrons to a Kaon and two same-flavor leptons. Likewise, the comparison of $e^-\mu^+$ to $e^+\mu^-$ in \cite{chris3} and \cite{epmum} is a test of CP symmetry in the lepton sector. Other symmetries could be used in similar 
implementations, such as forward-backward or time-reversal symmetries. 

Given the large number of symmetries in the SM which can be violated
in BSM scenarios, the potential benefits of implementing such symmetry-based
generic searches are significant. However, interpreting the results must be done with care. Indeed, a data-directed search will
naturally be tuned to identify regions including statistical fluctuations, or other measurement
effects which could induce asymmetries. If
a detected signal originates from a statistical fluctuation, it will disappear with more collected data. If it originates from a detector or other systematic effect which are correctly modeled in MC simulations then it can be ruled out. Any residual asymmetry can be considered a data-directed
BSM hypotheses, to be inspected using standard analysis techniques. In
this manner, the risk to claim a false discovery should not be higher than
when implementing hundreds of searches in the blind-paradigm, since
the trial factor is high in both cases \cite{eilamLook}.
% We argue that this
% is in fact an advantage of the method. If
% a detected signal originates from a statistical fluctuation, it will be
% washed out with more collected data. Signals which become more
% significant with accumulated data might be caused by some BSM
% process, but more likely will point to some detection effect which hasn't yet been
% correctly accounted for. This is a win-win scenario, as even in the
% absence of discovery, it can lead to a better understanding of
% the detector, and a more precise characterization of its response. We
% even argue that implementing general data validation tests, which
% verify if expected SM symmetries are indeed observed in the data,
% would have large benefits, first due to the discovery potential, but
% also in improving the overall data quality used in HEP searches, and
% should be standardly performed in this context.
\\

The aim of this paper is to draw the attention on the potential for discovering BSM physics when implementing searches in the DDP, and in particular, data-directed searches based on symmetries of the
SM. In this context, we lay the groundwork for a generic method to
compare  two data samples, and quantify the level of any discrepancy between
them, if present. As previously discussed, we do not address here the treatment of eventual systematic effects which can deteriorate the expected SM symmetry between the two samples. Nonetheless, as shown in \cite{emmeRun1} and \cite{epmum}, in analyses that were based on symmetry considerations, such effects can be accounted for.

Since the goal is to quickly scan multiple sub-regions of the observables space in a large number of final states, a fast method for identifying asymmetries is needed. We develop this method based on a simplified framework using MC simulated data. Different test statistics can be used to compare the two samples (e.g. Kolmogorov-Smirnov~\cite{ks-test}, student $t$-test~\cite{student-test}). In the implementation proposed in this paper, the samples are represented by 2D histograms\footnote{The generalization of the proposed analysis approach to $n$-dimensional histograms is straightforward.} of predetermined
properties of the data and compared using the simple $N_\sigma$ test statistic defined below. Since the method is fast, multiple 2D histograms
of all the existing properties and their combinations can be compared efficiently. We leave to future work the generalization of this study for a more comprehensive and optimized implementation.

When working with histograms, there is no a priori way to choose the bins, which is particularly challenging in many dimensions. One solution to this challenge is to make use of machine learning. Starting from~\cite{ano1,ano2} based on~\cite{ano3}, there have been a variety of proposals to perform anomaly detection with machine learning by comparing two samples~\cite{ano1,ano2,DAgnolo:2018cun,DAgnolo:2019vbw,Amram:2020ykb,gensearch2,Nachman:2020lpy,Andreassen:2020nkr,Benkendorfer:2020gek,Hallin:2021wme,dAgnolo:2021aun} (see Ref.~\cite{2010.14554,2112.03769,2102.02770,Kasieczka:2021xcg,Aarrestad:2021oeb} for recent reviews). Complementary to the binned DDP (henceforth, simply `the DDP'), we demonstrate that asymmetries can also be identified using weakly supervised Neural Networks (NN), similar to the approach in~\cite{ano3}. Nevertheless, for now such methods require training at least one NN for each event selection. This is time consuming and restricts the number of selections that can be tested, which could be limiting in the context of the DDP, depending on the available computational resources.

The sensitivity of the proposed DDP search is compared to that of two likelihood-based test statistics. While both assume exact knowledge of the signal shape, one represents an ideal search in which also the distribution of the symmetric background components are exactly known, and the other represents the expected sensitivity of a traditional blind analysis search employing a symmetry-based background estimation. According to the Neyman-Pearson lemma~\cite{neyman1933ix} these are the most sensitive tests for the respective scenarios they consider.

%In the following, we compare sensitivities obtained from three different test statistics. One is a simple generic test statistics that compares two measurements, which we use to demonstrate the symmetry-based DDP and refer to as the $N_\sigma$ test. It relies solely on the symmetry assumption and doesn't require  modeling the SM process or any potential signal.  The other two are likelihood-based test statistics which according to the Neyman-Pearson lemma (see e.g. \cite{nplemma}) are the most sensitive tests for the respective scenarios they consider. 

%The first ($q_0^{L1}$) needs full knowledge of the signal and assumes that the underlying distribution of the symmetric background component (the SM contribution) is perfectly known, therefore exhibiting the ideal sensitivity possible. The second ($q_0^{L2}$) follows the method used in the symmetry-based search for LFV Higgs decays \cite{emmeRun1}. It assumes full knowledge of the signal, but estimates the symmetric contribution of the SM processes from the two measurements. This represents the expected sensitivity of a traditional blind analysis search employing a symmetry-based background estimation. 
%the case where the only information is the symmetry, similar to the generic data-directed search proposed here. But since $q_0^{L2}$ has full knowledge of the signal, its sensitivity is the highest benchmark value that we can aim for. 

This paper is organized as follows.  Section~\ref{sec:quant} describes some of the statistical properties of the DDP symmetry search.  The simulated data used for our numerical studies is presented in Sec.~\ref{sec:data}.  Results for the DDP are given in Sec.~\ref{sec:results} and a complementary approach using neural networks is discussed in Sec.~\ref{sec:NN}.  The paper ends with conclusions and outlook in Sec.~\ref{sec:discussion}.

\section{Quantifying asymmetries}
\label{sec:quant}
Given two data samples, our goal is to determine the probability that they are \emph{asymmetric}, as opposed to originating from the same underlying distribution. The latter represents the null hypothesis, where both measurements are indeed symmetric as expected from the symmetry property of the SM considered. In the context of the symmetry-based DDP proposed here, and unlike in other statistical tests commonly used in
BSM searches, no signal assumptions are made. The test is intended to output the
probability at which the background-only hypothesis is rejected.

%Any deviation
%from the expected symmetry is a discrepancy which should be further
%investigated. As a result, there isn't an alternative
%hypothesis to test against, rather the test should output some
%probability at which the background-only hypothesis is rejected. 

In order to rapidly scan many selections and final states, the method used to quantify the asymmetry between two samples should be efficient. This can be achieved if we ensure that the results obtained are independent of the properties of the underlying symmetric background component. Indeed, one of the most time consuming tasks for implementing a statistical test to reject an hypotheses is the determination of the test statistic's probability distribution function (PDF) under said hypotheses. But if this PDF is constant and known, we avoid the time consuming task of deriving it for each different samples tested.

The generic $N_{\sigma}$ test statistic considered is given in
Equation \ref{eq:nsigma}. $A$ and $B$ are two $n$-dimensional matrices, representing the two tested data samples projected into histograms of $n$ properties of the measurements. They each have $M$ bins in total, the $A_i$ and $B_i$ are their respective number of entries in bin $i$, and the $\sigma_{Ai}$ and $\sigma_{Bi}$ their respective standard errors:
\begin{equation}
  \label{eq:nsigma}
  \mathrm{N}_{\sigma}(B,A) = \frac{1}{\sqrt{M}} \sum_{i=1}^{M} \frac{B_i - A_i}{\sqrt{\sigma_{Ai}^2+\sigma_{Bi}^2}}\,.
\end{equation}
In this formalism, we search for a signal in $B$ by comparing it to
the reference measurement $A$, but their roles 
are exchangeable. When $A$ and $B$ are two (Poisson-distributed)
measurements, Equation \ref{eq:nsigma} simplifies to: 
\begin{equation}
  \mathrm{N}_\sigma(B,A) = \frac{1}{\sqrt{M}} \sum_{i=1}^{M}
  \frac{B_i - A_i}{\sqrt{A_i+B_i}}\,.
  \label{eq:nsigma2}
\end{equation}
It can be shown that in the limits of
the normal approximation, applicable here provided there are enough
statistics in each bin of the two matrices, the symmetry-case PDF of
the $N_{\sigma}$ test is well
approximated by a standard Gaussian.
This satisfies the condition that the test should be independent of
the underlying symmetric component, ensuring its efficiency. In what
follows, we confirmed that this approximation is valid when ensuring at
least 25 entries per bin. For scenarios with lower statistics, the
distortion of the background-only PDF from the normal distribution
should be evaluated. Nevertheless, large $N_{\sigma}$ values would
correspond to asymmetries. 

The performance of the $N_{\sigma}$ test is compared to that of two distinct likelihood-based test statistics, which are built on the test statistic for discovery of a positive signal introduced in \cite{eilamStat} and rely on the full knowledge of the signal shape that is being searched for:
% This
% likelihood-based test gives
% the maximum power to reject the background-only hypothesis, in the
% case where the alternative hypothesis, or background + signal
% hypothesis, is given. Meaning that this test is
% not generic, but rather used to search for specific, predefined, signal. Being the most sensitive test, it is appropriate for setting a reference benchmark for the performance of the method developed in this study. 
% We consider two scenarios for comparison:
\begin{itemize}
    \item $q_{0}^{L1}$ assumes that the underlying symmetric component is perfectly known. This is equivalent to the ideal analysis case in which the signal and background distributions are perfectly known (no uncertainties).
    \item $q_{0}^{L2}$ uses no a priori knowledge of the underlying symmetric distribution, and estimates it from the two measurements as part of the fitting procedure. This represents the case where the symmetry is the only available information.
\end{itemize}
Since we aim at comparing the sensitivity to detect asymmetries using the
$N_\sigma$ test relatively to the
likelihood-based tests, statistical uncertainties on the signal are not included in this study. The likelihood functions for each scenario are
shown below, where $S$ is the shape of the signal considered, $B$ is the tested sample, $T$ is the true distribution of the symmetric background and $A$ is a measurement of $T$. The parameter $\mu$ represents the signal-strength, and $b=\{b_i\}$ are the background parameters (one per bin of the matrix):
\begin{align}
  \label{eq:likelihood}
    &L1_{\mu}(B,T,S) = \mathrm{Poisson}(B~|~T+\mu S)\\
    &L2_{\mu}(B,A,S;b) = \mathrm{Poisson}(B~|~b+\mu S) \cdot \mathrm{Poisson}(A~|~b)
\end{align}
%  L1_{\mu}(B,T,S) &=& \mathrm{Poisson}(B~|~T+\mu S) \\
%  &=& \prod_{i=1}^{M}\frac{(T_i+\mu S_i)^{B_i}}{B_i!}e^{-(T_i+\mu S_i)}
%\end{equation}
%\begin{equation}
%  \label{eq:likelihood2}
%  L2_{\mu}(B,A,S;b) &=& \mathrm{Poisson}(B~|~b+\muS)\cdot\mathrm{Poisson}(A~|~b) \\
%                   &=& \prod_{i=1}^{M}\frac{(b_i+\mu S_i)^{B_i}}{B_i!}e^{-(b_i+\mu S_i)}\frac{(b_i)^{A_i}}{A_i!}e^{-b_i}
The formalism used, which permits a  comparison with the $N_\sigma$ test, is shown in
Equations \ref{eq:qtest1} and \ref{eq:qtest2}, where $L_\mu$ is the
likelihood function (either $L1_\mu$ or $L2_\mu$), $\lambda_\mu$ is the profile likelihood ratio,
$\hat{\mu}$ and $\hat{b}$ are the maximum likelihood
estimators of $\mu$ and the $b_i$ parameters, and $\hat{\hat{b}}$ is the maximum likelihood
estimator of the $b_i$ when $\mu$ is fixed. 
\begin{align}
  &\lambda_\mu(B,A,S)=\frac{L_\mu(B,A,S;\hat{\hat{b}})}{L_{\hat{\mu}}(B,A,S;\hat{b})}   \label{eq:qtest1}  \\
  &q_0(B,A,S) = \left\{
    \begin{array}{ll}
      -2\ln\lambda_0(B,A,S) &, \hat{\mu} \geq 0 \\
     +2\ln\lambda_0(B,A,S) &, \hat{\mu} < 0
    \end{array} \right. \label{eq:qtest2}
\end{align} 
% The two-sided $q_0$ test defined in Equation \ref{eq:qtest2} is similar
% to the one introduced in \cite{eilamStat}. 
%This formalism permits a straightforward comparison to the $N_\sigma$ test; 
% For each scenario, it simplifies to (showing only when
% $\hat{\mu}\geq0$)\footnote{Here we use the fact that $\hat{\hat{b}}$,
%   the MLE of $b$ when $\mu=0$, is known and equal to the average of the
%   two matrices A and B}:
% \begin{align}
%   \label{eq:q01} 
%   &q_{0}^{L1,~\hat{\mu}\geq0}(B,T,S) = 2(-\hat{\mu} +\sum\limits_{i=1}^{M}[B_i\ln(1+\hat{\mu}\frac{S_i}{T_i})])
% \end{align}
% \begin{multline}
%   \label{eq:q02}
%   q_{0}^{L2,~\hat{\mu}\geq0}(B,A,S) = 2(-\hat{\mu} + \sum\limits_{i=1}^{M}[(A_i+B_i)-2\hat{b_i}\\
%   +B_i\ln(\frac{2(\hat{b_i}+\hat{\mu}S_i)}{A_i+B_i})+A_i\ln(\frac{2\hat{b_i}}{A_i+B_i})])
% \end{multline}
When performing a test for discovery, we compare the test's score to the
background-only PDF to obtain a $p$-value ($p$) which gives a
measure of the level at which the background
hypothesis can be rejected. We then translate this $p$-value
into an equivalent significance $Z =\Phi^{-1}(1-p)$, where $\Phi^{-1}$
is the quantile of the standard Gaussian. A
significance of 5 is commonly considered an appropriate level to constitute a
discovery, corresponding to $p\approx2.87\times 10^{-7}$. For the case of
the $N_\sigma$ test, the background-only PDF is itself a standard
Gaussian. Therefore the score obtained is directly a measure of the
obtained significance $Z$, bypassing the need to compute the
$p$-value:
\begin{equation}
  Z=N_\sigma(B,A)\,.
\end{equation}
Similarly, regarding the $q_0$ test, we know from \cite{eilamStat} that:
\begin{equation}
  Z=\sqrt{q_0}(B,A,S)\,.
\end{equation}
So the $\sqrt{q_0}$ background-only PDF is again a standard Gaussian\footnote{This can
  also be shown in the more common single-sided formalism presented in \cite{eilamStat}, where the background-only PDF of $q_0$ in the asymptotic limit is given by
$\frac{1}{2}(\delta(0)+\chi_1^2)$ where $\chi_1^2$ is the $\chi^2$ distribution with one degree of freedom. Thus the PDF of $\sqrt{q_0}$ is 
$\frac{1}{2}(\delta(0)+\chi_1)$, and the $\chi_1$ distribution is the
\emph{half-normal} distribution.}. Therefore, in the following, we directly compare the $N_\sigma$ and $\sqrt{q_0}$ significance values.

\section{Data preparation}
\label{sec:data}
\begin{figure}[ht]
  \centering 
  \begin{subfigure}{\columnwidth} 
    \centering
    \includegraphics[width=.99\linewidth]{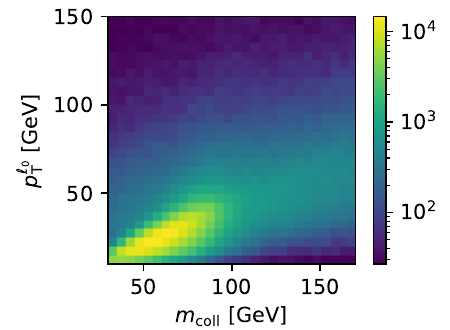}
  \end{subfigure}
  \begin{subfigure}{\columnwidth}
    \centering
    \includegraphics[width=.99\linewidth]{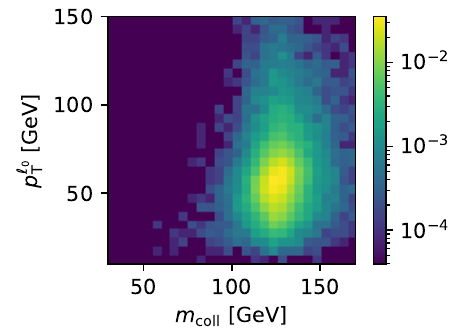}
  \end{subfigure}
  \caption{The $e/\mu$ background template matrix $T$ (top) and
    the Higgs LFV signal template matrix $S$ (bottom). The $x$, $y$ and $z$ axes are the
    collinear mass, leading lepton $p_\text{T}$ and number of entries per bin respectively ($S$ is normalized).}
  \label{fig:hlfv_tpls}
\end{figure}

The symmetry-based DDP is demonstrated in a
practical example, the search for Higgs LFV decays, $H\rightarrow\tau\mu$ where the $\tau$ further decays to an electron. The  SM processes considered which contribute to the symmetric background includes Drell-Yan, di-boson, $Wt$, $t\bar{t}$ and SM Higgs ($H\rightarrow WW / \tau\tau$). 
For each of these processes, a sample equivalent to $40~\mathrm{fb}^{-1}$ of $pp$ collisions at $\sqrt{s} = 13$ TeV was generated using MadGraph~2.6.4~\cite{madgraph} and Pythia~8.2~\cite{pythia}. The response of the ATLAS detector was emulated using Delphes~3~\cite{delphes}.  
The signal processes considered are gluon-gluon fusion and vector
boson fusion Higgs production mechanisms.
These SM events are used to construct an ${e\mu}$ symmetric template
($T$) matrix -- representing the SM background underlying distributions
from which symmetric samples will be drawn (see description of this process further below). The
Higgs LFV signal events are used to construct a normalized 
signal template matrix $S$. This is done by projecting
the simulated measured events on a $28\times28$ 2D histogram, with two
selected event properties:
\begin{itemize}
\item x-axis: collinear mass (defined e.g. in \cite{emmeRun1}), 5 GeV bins from 30-170 GeV
\item y-axis: leading lepton $p_\text{T}$, 5 GeV bins from 10-140 GeV
\end{itemize}
% The different SM processes used to constitute $T$ are
% $Z/\gamma^*\rightarrow\ell\ell$, $Z/\gamma^*\rightarrow\tau\tau$,
% $WW$, $Wt$, $t\bar{t}$, $(ggF + VBF)H \rightarrow WW$, and $(ggF +
% VBF)H \rightarrow\tau\tau$.
% The signal processes which constitute $S$ are $(ggF +
% VBF)H \rightarrow\mu\tau$.
To demonstrate the concept, and to allow quantitative comparisons to
the performance of the likelihood-based tests, we avoided bins with
low statistics by adding a flat 25 entries to each bin 
in $T$. The resulting $T$ and $S$ templates are shown in Figure
\ref{fig:hlfv_tpls}.

Other background and signals considered are flat $T$ background
distributions (with either 100 or $10^4$ entries per bin), and rectangle
and 2D Gaussian signals $S$ templates. 
%, which are described further, in an attempt to further characterize the performance of the $N_\sigma$ test.

Given a background template $T$, which represents the
underlying symmetric distribution, and a signal template $S$,
which can be injected with different levels of signal-strength, the
procedure to generate the samples used to qualify the different tests
is as follows. From $T$ we Poisson draw $N$ pairs of $(A, B)$
background-only measurements which are symmetric up to statistical
fluctuations. The background + signal measurements $B^s$ are obtained
by injecting some signal into the $B$ samples. We inject the signal with a
signal-strength $\mu_\mathrm{inj}$, determined such that a $q_0$
test for discovery ($q_0^{L1}$ or $q_0^{L2}$) outputs a given significance $Z_\mathrm{inj}$ when testing
${B^s=B+\mu_{\mathrm{inj}}S}$ against $B$:
\begin{equation}
  \label{eq:q0z}
  \sqrt{q_0}(B+\mu_{\mathrm{inj}} S, B, S)=Z_{\mathrm{inj}}\,.
\end{equation}
Since $S$ is normalized, $\mu_{\mathrm{inj}}$  is the number of signal
events added to the $B$ sample. 

Explicitly, for the $q_0^{L1}$ and $q_0^{L2}$ cases, it is 
found by solving Equations \ref{eq:muinj1} and \ref{eq:muinj2}, respectively:
\begin{align}
  \label{eq:muinj1}
  2\left(-\mu_{\mathrm{inj1}} + \sum\limits_{i=1}^{M}\left[
  (B_i+\mu_{\mathrm{inj1}}S_i)\ln\left(1+\mu_{\mathrm{inj1}}\frac{S_i}{B_i}\right)\right]\right) = Z_{\mathrm{inj1}}^2
\end{align}
\begin{multline}
  \label{eq:muinj2}
  2\sum\limits_{i=1}^{M}\left[(B_i+\mu_{\mathrm{inj2}} 
  S_i)\ln\left(1+\mu_{\mathrm{inj2}}\frac{S_i}{2B_i+\mu_{\mathrm{inj2}} S_i}\right)\right.\\
  \left.-B_i\ln\left(1+\mu_{\mathrm{inj2}}\frac{S_i}{2B_i}\right)\right]=Z_{\mathrm{inj2}}^2
\end{multline}

For each separate experiment considered and detailed below, the number of
$A$, $B$ and ${B^s=B+\mu_{\mathrm{inj}}S}$ matrices we generate is
$N=20000$. For the $N_{\sigma}$ and
$q_{0}^{L2}$ tests, the PDFs of the symmetric case (background-only)
are obtained by comparing the $B$ and $A$ pairs, and the PDFs of the
asymmetric case (signal+background) by comparing the $B^s$ and $A$
pairs. The same is applied for the $q_{0}^{L1}$ test, when the $A$ matrices are
replaced by the template $T$.

% \bibliographystyle{spphys}
% \bibliography{symsearch}

%\tableofcontents

\section{Results}
\label{sec:results}
\begin{figure}[ht]
  \centering
    \includegraphics[width=\columnwidth]{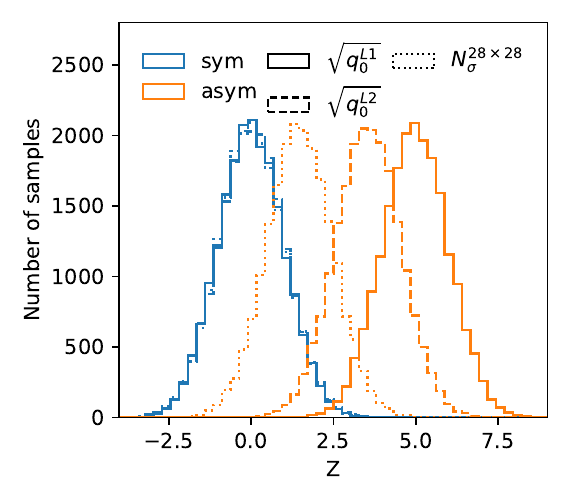}
    \caption{Significance PDFs comparing results of the $N_\sigma$,  $q_{0}^{L1}$ and $q_{0}^{L2}$ tests for the Higgs LFV example, with injected signal strength corresponding to $5\sigma$ of $q_{0}^{L1}$.}
    \label{fig:pdfs_hlfv} 
\end{figure}

Focusing on the Higgs LFV example, using the signal ($S$) and background ($T$)
templates shown in Figure \ref{fig:hlfv_tpls}, we apply an
injected signal-strength $\mu_\mathrm{inj}$ which corresponds to
$5\sigma$ significance of the ideal $q_{0}^{L1}$ test. To give an impression, when applied to $T$, this corresponds to a signal fraction of 0.2\%, or in a $6\times6$ window centered around the signal of 2.8\%.
In Figure \ref{fig:pdfs_hlfv}, we compare $Z$ PDFs obtained with the
$q_{0}^{L1}$, $q_{0}^{L2}$ and $N_\sigma$ tests. As expected, the
symmetric-case PDFs of all tests are consistent with standard
Gaussian distributions. We observe that the
background + signal (asymmetric-case) PDFs are consistent with Gaussians with variance $1\pm0.05$ (for all examples considered), centered around the
resulting average significance $Z_\mathrm{avg}$ of the relevant test. The $Z_\mathrm{avg}$ of each test can be directly estimated
by using the \emph{Asimov} data \cite{eilamStat}; setting $A=T$ and
$B^s=T+\mu_\mathrm{inj}S$.
The resulting significance with the $q_{0}^{L1}$ test is predictably
$Z_{\mathrm{avg}}=5.0\approx Z_{\mathrm{inj}}$. With $Z_{\mathrm{avg}}=3.53$, $q_{0}^{L2}$ is less sensitive than $q_{0}^{L1}$ since it
does not use an a priori knowledge of the background, but estimates it from the two measurements as part of the fitting procedure.
Since the $N_\sigma$ test is averaged on all the bins, and most of them only include background contributions, the resulting average significance $Z_\mathrm{avg}=1.48$ is significantly lower than the separation power measured with the $q_{0}^{L2}$ test.

\begin{figure}[ht]
    \centering
    \includegraphics[width=\columnwidth]{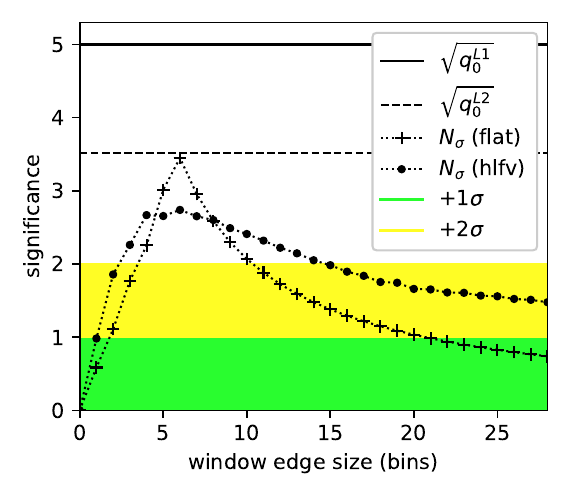}
  \caption{Significance measured from the Asimov data, with the $N_\sigma$ test applied to increasing window sizes, and compared to the $q_{0}^{L1}$ and $q_{0}^{L2}$ significance. Results for the Higgs LFV example and the ideal (flat) scenario are shown, with injected signal strength corresponding to $5\sigma$ of $q_{0}^{L1}$. The green and yellow bands correspond to the 1$\sigma$ and $2\sigma$ deviations from the symmetry (no signal) assumption, respectively.}
  \label{fig:zscan_wsize} 
\end{figure}

In general, it can be much more efficient to apply the $N_\sigma$ test in a sub-region of the data samples. Even though the signal's shape and location is not known in a generic test, since the calculation of $N_\sigma$ is fast, one
could test multiple bin subsets\footnote{There is a trials factor for performing multiple tests, but as stated earlier, the goal is to identify interesting regions and not to compute a precise global $p$-value.  That could be done with $k$-folding or other divide-and-test schemes, which we leave for future work to explore.}, or develop an algorithm
to optimize this selection. In Figure \ref{fig:zscan_wsize}, we show $N_\sigma$ scores with the Asimov data, obtained when the test is performed on square windows of different sizes, centered around the location of the signal. The
$N_\sigma$ sensitivity increases when the window encapsulates the signal region more precisely, reaching up to
$Z_\mathrm{avg,max}=2.74$ with the $6\times6$ bins window. Thus, for this example, the sensitivity achieved is only slightly worse
than the one achieved with the $q_{0}^{L2}$ test, which exploits a full knowledge of the signal shape. The $N_\sigma$ results presented hereafter are for the best suited window ($6\times6$ bins for all examples considered).

\begin{figure}[ht]
  \centering
    \includegraphics[width=\columnwidth]{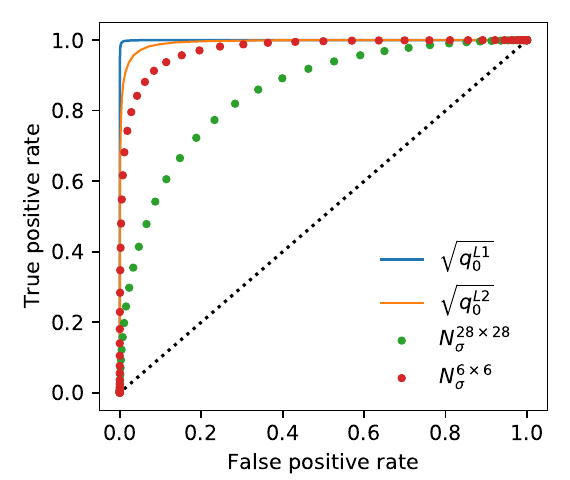}
  \caption{ROC curves comparing results of the $N_\sigma$, $q_{0}^{L1}$ and $q_{0}^{L2}$ tests for the Higgs LFV example, with injected signal strength corresponding to $5\sigma$ of $q_{0}^{L1}$.}
    \label{fig:rocs_hlfv} 
\end{figure}

\begin{figure}[ht]
    \centering
    \includegraphics[width=\columnwidth]{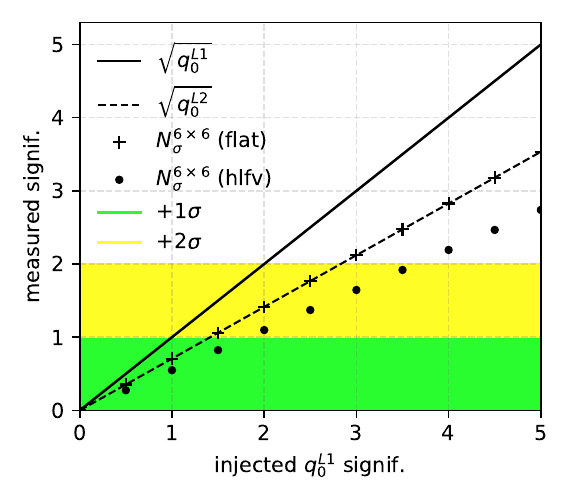} 
  \caption{Significance measured from the Asimov data for increasing injected signal, comparing results of the $N_\sigma$, $q_{0}^{L1}$ and $q_{0}^{L2}$ tests. Results for the Higgs LFV example and the ideal (flat) scenario are shown. The green and yellow bands correspond to the 1$\sigma$ and $2\sigma$ deviations from the symmetry (no signal) assumption, respectively.}
      \label{fig:zscan_zinj} 
\end{figure}

In Figure \ref{fig:rocs_hlfv} we show the Receiver Operating Characteristic (ROC) curves obtained from the PDFs of the different tests. The Area-Under-Curve (AUC) measured is approximately 1.0 for the $q_{0}^{L1}$ test and 0.994 for the $q_{0}^{L2}$ test. With an AUC of $=0.973$, the $N_\sigma$ test is only 2.6\% less sensitive than the $q_{0}^{L1}$ test, and 2.0\% less sensitive than the $q_{0}^{L2}$ test. Finally, in Figure \ref{fig:zscan_zinj}, we show $Z_\mathrm{avg}$ per test (estimated from the Asimov data), for increasing injected signal strength. Using the $N_\sigma$ test statistic, the symmetric case (background only) can be separated from the asymmetric case at the level of $2\sigma$ if the signal that would have been measured assuming an ideal analysis ($q_0^{L1}$) is at the level of $3.5\sigma$. This should be compared also to the $2.5\sigma$ separation that would have been obtained in the same case using the profile likelihood ratio test statistic that uses the two samples to estimate the symmetric background and a full knowledge of the signal shape ($q_0^{L2}$).

For clarity, we also consider a \emph{flat} background template $T$ with $10^4$ entries in each bin, and a \emph{flat rectangle} signal template $S$ of size $6\times6$ bins, located at the center of $T$. Since the $q_{0}^{L1}$ and $q_{0}^{L2}$ are independent of the background and signal shapes, and only depend on the injected signal strength, their symmetry- and asymmetry-case PDF will remain unchanged. The PDF associated with the $N_\sigma$ in the asymmetric case will change. As shown in Figures \ref{fig:zscan_wsize} and \ref{fig:zscan_zinj}, in this simplified case the $N_\sigma$ sensitivity matches exactly the sensitivity of $q_{0}^{L2}$ test. This hints that the loss of sensitivity of the generic $N_\sigma$ test, compared to $q_{0}^{L2}$, is mainly due to shape variations of the background and the signal (in the optimal sub-region that is tested). But even in a realistic scenario like the Higgs LFV example, the sensitivity loss is reasonable (from $Z_\mathrm{avg}=3.53$ to 2.74) and the power achieved to identify regions with asymmetry, even though the $N_\sigma$ test is generic, is significant.
%For the other shapes of signal and background considered, includes \emph{aaa,bbb,cc}.
In terms of the ability to identify asymmetries, similar performance was obtained for all the other shapes of signal and background considered.

%In addition to the results presented here, we also considered different types of background and signal templates. Some additional results are shown in the appendix, but in summary the behavior is similar to what was shown here.  
% This is enough power to identify regions of interest,
% even though the $N_\sigma$ test is generic.

\section{Identifying asymmetries with Neural Networks}
\label{sec:NN}
%Something about the signal stat uncertainty here can't be ignored

Machine learning-based anomaly detection methods constructed by comparing two samples are categorized as weakly- or semi-supervised learning because both samples are mostly background and one of them will have more signal than the other. The sample with more potential signal is given a noisy label of one and the other sample is given a label of zero.  A classifier trained to distinguish the two samples can then automatically identify subtle differences between the samples without explicitly setting up bins.  Existing proposals construct the samples from signal region / sideband regions \cite{ano1,ano2,gensearch2}, from data versus simulation \cite{DAgnolo:2018cun,DAgnolo:2019vbw,dAgnolo:2021aun,Chakravarti:2021svb}, as well as other approaches~\cite{Amram:2020ykb,Nachman:2020lpy,Andreassen:2020nkr,Benkendorfer:2020gek,Hallin:2021wme}.  We propose to extend this methodology to symmetries.  

The combination of machine learning and symmetry has received significant attention.  For a given symmetry, one can construct machine learning methods that are invariant or covariant (in machine learning, this is called \textit{equivariant}) under the action of that symmetry.  For example, recent proposals have shown how to construct Lorentz covariant neural networks \cite{symNN1,symNN2,Qiu:2022xvr}. Symmetries can also be used to build a learned representation of a sample \cite{symNN3}. There have also been proposals to use machine learning methods to discover symmetries automatically in samples \cite{symNN4,symNN5,symNN6}. In the context of BSM searches, Ref.~\cite{chris1,chris2} recently described how to use a weakly supervised-like approach to test if a given symmetry is broken by applying the transformation to the input data. Our approach also starts by positing a symmetry, but we do not apply the symmetry transformation to each data point. Instead, we have two samples which should be statistically identical in the presence of a symmetry, but which could be different when BSM is present.

\begin{figure}[ht]
  \centering
    \includegraphics[width=\columnwidth]{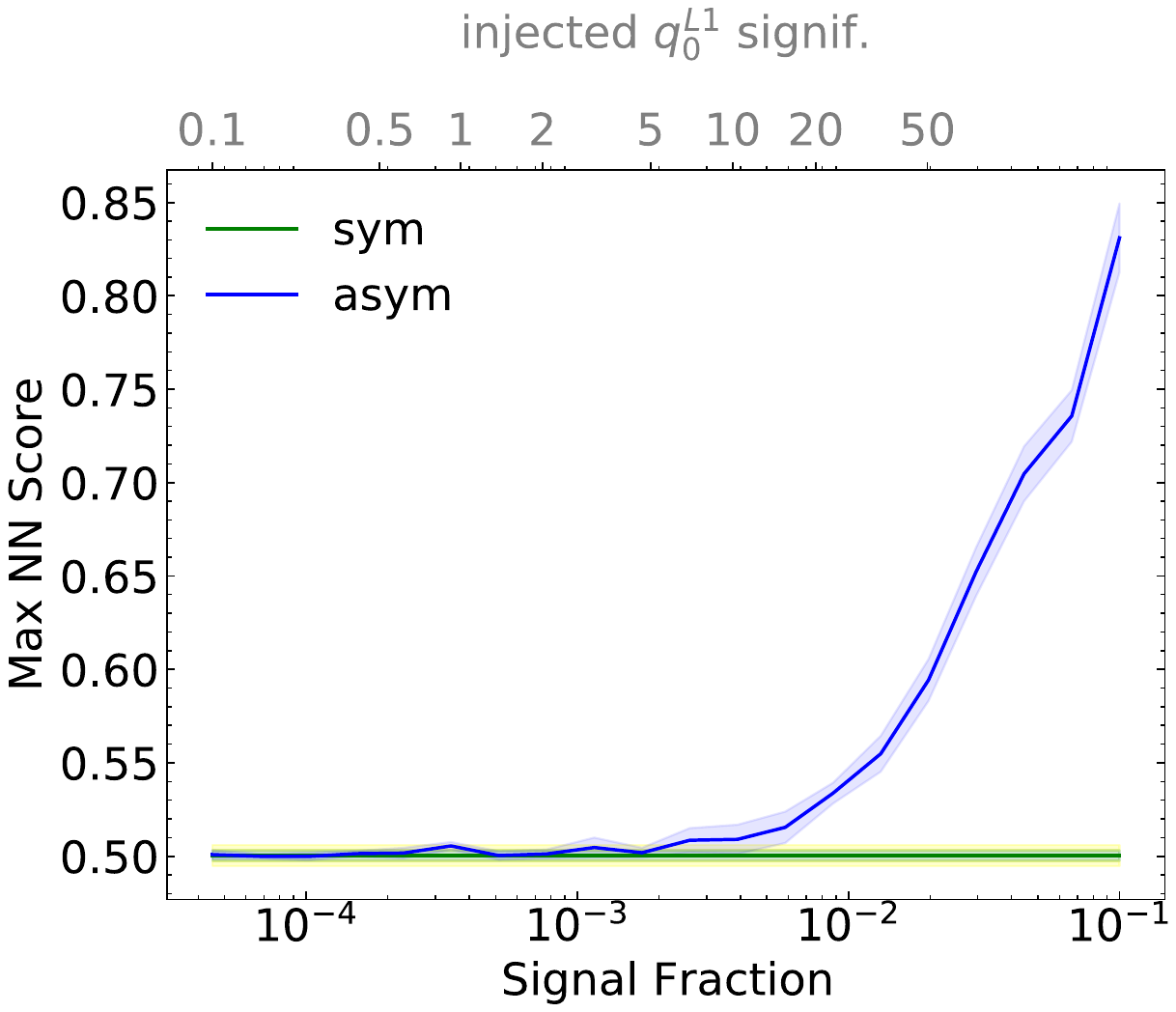}
  \caption{The maximum neural network score from training a classifier to distinguish the $e\mu$ from $\mu e$ samples with (asym) and without (sym) a BSM contribution.  The green (yellow) and blue bands represent (twice) the standard deviation over 10 bootstrap samples.The separation power is shown as a function of the injected signal fraction (bottom scale) and the corresponding significance calculated with the ideal $q_0^{L1}$ test. Note that these results are not directly comparable to the binned DDP because it is not possible to ignore signal statistical uncertainties.}
    \label{fig:NN} 
\end{figure}

In the following, we demonstrate the concept of identifying asymmetries using a weakly supervised approach. Considering the $e\mu$ symmetry example discussed above, one of the samples is the $e\mu$ sample and the other is the $\mu e$ sample. The same two-dimensional space as described earlier is used for illustration; extending to higher dimensions is technically straightforward.  A deep neural network with three hidden layers and 50 nodes per layer is used for the classifier.  Rectified Linear Units (ReLU) are used for all intermediate layers and the output is passed through a sigmoid function.  This network is implemented using Keras~\cite{keras} and Tensorflow~\cite{tensorflow} using Adam~\cite{adam} for optimization.  We train for 20 epochs with a batch size of 200.  None of these parameters were optimized. Figure \ref{fig:NN} shows the symmetry/asymmetry separation power of the NN as a function of the signal fraction injected to the $\mu e$ sample.  The background-only band is computed via bootstrapping~\cite{efron1979}.  For each bootstrap, two samples are created by drawing from the $e\mu$ and $\mu e$ events with replacement.  By mixing the two samples, any asymmetry is removed.

There is no unique way to quantify the NN performance. An optimal test statistic by the Neyman-Pearson Lemma~\cite{neyman1933ix} is monotonically related to the likelihood ratio.  Ref.~\cite{DAgnolo:2018cun,DAgnolo:2019vbw,Nachman:2021yvi,dAgnolo:2021aun} show how to modify the loss function so that the average loss approximates the (log) likelihood ratio.  Here, we find that in practice, the maximum NN score using the standard binary cross entropy loss function is an effective statistic, which goes from 0.5 in the case of no signal and increases as more signal is injected.  The background-only band in Figure \ref{fig:NN} is computed via bootstrapping and where the blue line and green/yellow bands cross indicate the approximate $1\sigma/2\sigma$ exclusion. The NN is able to automatically identify the presence of BSM for signal fractions that are a few per mil, corresponding to around 5$\sigma$ significance calculated with the ideal $q_0^{L1}$ test.  Future explorations of this idea will understand the best way to set up the training, what statistics are most effective, and how to best extend to higher dimensions.

%\bibliographystyle{spphys}
%\bibliography{symsearch}

%\tableofcontents
\section{Discussion}
\label{sec:discussion}
%BACK TO DDP
%MORE IMPLEMENTATIONS
%- statistical analysis: outline
%- compare matrices: simplified scheme, easily generalizable
%- the nsigma test: generic, simple, efficient
%- q0 - reference: gen data + comp results
%- result interpretation
%- generated data: T and S templates, MC simul, Zinj(q0)
%- results: simple example, varying templates
%- results: HLFV
%- known background scenario

With limited resources at hand and yet no conclusive indication of BSM physics found, we must try novel and complementary avenues for discovery. To overcome the limitations stemming from adapting the blind-analysis strategy, we  propose developing the DDP. Similarly to \cite{D0:2000vuh,H1:2008aak,gensearchCDF,gensearch1,gensearch3} yet without relying on MC simulations, its principal objective is to allow scanning as many regions of the observable-space as possible and direct dedicated analyses towards the ones in which the data itself exhibits deviation from some fundamental and theoretically well-established property of the SM. Relative to regions in which the data agrees well with the SM predictions, the ones that exhibit deviations are promising for further investigations into BSM physics.  

We propose developing the DDP based on symmetries of the SM and demonstrate its potential sensitivity using as an example the $e/\mu$ symmetry. Symmetries allow splitting the data into two mutually exclusive samples which, under the symmetry assumption, differ only by statistical fluctuations. Thus, asymmetry observed between the two samples in any observable and at any sub-selection of these samples, is potentially interesting and should be considered for further study. 

While different algorithms can be developed to identify asymmetries, even the most simple one developed, the $N_\sigma$ test statistic, already provides good sensitivity. It is compared to the sensitivity obtained with two likelihood-based test statistics; the first, $q_0^{L1}$, represents an ideal analysis in which both the signal and the symmetric contribution from the SM processes are perfectly known. The second, $q_0^{L2}$, represents the expected sensitivity of a traditional blind analysis search for a predefined signal that employs a symmetry-based background estimation (\cite{emmeRun1}). 

Compared to the sensitivity obtained in an ideal analysis, the separation power between the symmetric case and an asymmetry at the level of 5$\sigma$ is less than 3\% lower in terms of the area under the ROC curve, and a separation at the level of $2\sigma$ is achieved for $3.5\sigma$ signal injected. Compared to a traditional symmetry-based analysis, the separation power between the symmetric case and an asymmetry at the level of 3.5$\sigma$ is less than 2\% lower in terms of the area under the ROC curve, and a separation at the level of $2\sigma$ achieved using the $N_\sigma$ test is only slightly degraded relative to the $2.5\sigma$ obtained with the $q_0^{L2}$ test. The results quoted are when applying the $N_\sigma$ test in the best suited window for the examples considered. The ability to find this optimal window demonstrates the strength of the DDP. Since the test is rapid, a large number of n-dimensional histograms and windows within  can be tested efficiently. This could permit scanning the data systematically in search for asymmetries.

We have shown that weakly-supervised NNs can also be used to identify asymmetries between two samples. This paves the way towards NN based DDP. 

We emphasize that traditional blind-analyses are expected to be the most sensitive ones for any predefined signal. Nonetheless, it is impossible to conduct a dedicated search in any possible final state and at any possible event selection. Moreover, not all potential signals can be thought of. Thus, the DDP could significantly expand our discovery reach.

\section*{Acknowledgements}
SB is supported by grants from the Israel Science Foundation (grant
number 2871/19), the German Israeli Foundation (grant number
I-1506-303.7/2019) and by the Yeda-Sela (YeS) Center for Basic
Research. BN is supported by the U.S. Department of Energy (DOE), Office of
Science under contract DE-AC02-05CH11231. 

\section*{Data Availability Statement}
The data that support the findings of this study are available from
the corresponding author, MB, upon reasonable request.

\bibliographystyle{spphys}
\bibliography{main}
\end{document}